\documentclass[
    ,final            
  ]
  {aipproc}

\layoutstyle{6x9}

\usepackage{epsf,graphics,amssymb,amsfonts,amsmath,array}

\newcommand{\lQ}{\Lambda_{\rm QCD}}

\newcommand{\als}{\alpha_{\rm s}}

\newcommand{\siml}{{\ \lower-1.2pt\vbox{\hbox{\rlap{$<$}\lower6pt\vbox{\hbox{$\sim$}}}}\ }} 
\newcommand{\simg}{{\ \lower-1.2pt\vbox{\hbox{\rlap{$>$}\lower6pt\vbox{\hbox{$\sim$}}}}\ }}

\begin{document}

\title{Quarkonium in a weakly-coupled quark-gluon plasma}


\author{Antonio Vairo}{
address={Physik-Department, Technische Universit\"at M\"unchen,
James-Franck-Str. 1, 85748 Garching, Germany}
}

\begin{abstract}
We report about a recent calculation of the heavy quarkonium mass and decay width 
in a quark-gluon plasma, whose temperature $T$ and screening mass $m_D$ satisfy the hierarchy 
$m\als \gg \pi T \gg m \als^2 \gg m_D$, $m$ being the heavy-quark mass, up to order $m\als^5$.
The calculation may be relevant to understand the behavior of the $\Upsilon(1S)$ 
in a quark-gluon plasma at present-day colliders.
\end{abstract}

\maketitle

\section{Introduction}
The behavior of quarkonium in a medium has been the object of many studies 
since it was realized that the medium may induce quarkonium dissociation 
and that the dissociation pattern of the different quarkonia may provide an effective 
thermometer of the medium~\cite{Matsui:1986dk}.
How the medium influences the quarkonium and eventually 
leads to its dissociation is the problem that such studies address.

In recent years, new insight in the problem has been gained by studying 
low-lying quarkonia in a weakly-coupled quark-gluon plasma. 
Low-lying quarkonia, like the bottomonium and charmonium 
ground states, are (believed to be) characterized by a typical inverse radius, which is of the order of  
$m\als$, larger than the non-perturbative scale of QCD, $\lQ$, and 
a typical energy $E$ of the quarks in the bound state,  which is of the order of  
$m\als^2$, larger than or of the same order as $\lQ$.  
Such states are Coulombic. In the case of the bottomonium ground state, 
it is often also assumed that $E \gg \lQ$.
In a weakly-coupled plasma, both the temperature $T$ of the plasma and the  
screening mass, $m_D$, are larger than $\lQ$. Moreover, the perturbative power counting 
implies that $m_D$ is suppressed with respect to $T$, for $m_D^2$ shows up at one loop.
With these specifications, observables describing low-lying quarkonia in a weakly-coupled quark-gluon plasma 
may be calculated to a large extent analytically in perturbation theory, which 
makes the interest of these systems. As we will argue later, this could be the case of 
the $\Upsilon(1S)$ at the LHC.

In \cite{Laine:2006ns,Laine:2007gj,Laine:2007qy,Burnier:2007qm}, the perturbative quarkonium 
static potential was studied for quark-antiquark distances $r$ such that $T \gg 1/r \simg m_D$.
The surprising output of these studies has been that the imaginary part of the gluon self energy, 
which may be traced back to the Landau-damping phenomenon, induces an  imaginary part of the static 
potential and hence a thermal width of the static quark-antiquark bound state.
In \cite{Beraudo:2007ky}, static particles in a hot QED plasma were 
considered in the situation $1/r \sim m_D$, confirming previous results.
In \cite{Brambilla:2008cx}, the quarkonium static potential was first studied in an effective 
field theory (EFT) framework that exploits systematically the hierarchy of different energy scales in the 
problem. The static potential was studied for distances that range from larger to smaller than $1/T$  
and a new dissociation mechanism was identified. This is the color-singlet to color-octet break-up 
mechanism that provides the dominant contribution to the thermal decay width when $E \gg m_D$.
In \cite{Brambilla:2010xn}, moreover, the relation between the quarkonium potential and the 
correlation of two Polyakov loops, a quantity often evaluated on the lattice, has been investigated. 
A comprehensive study of non-relativistic bound states in a hot QED 
plasma in a non-relativistic EFT framework was performed in \cite{Escobedo:2008sy,Escobedo:2010tu}.
It has been pointed out that quarkonium melts at temperatures larger than the melting temperature, $T_{\rm melting}$, 
defined as the temperature for which the thermal decay width becomes as large as the binding energy; 
parametrically, it holds that $\pi T_{\rm melting} \sim m g^{4/3}$.
Since this temperature is lower than the temperature at which screening sets in, 
i.e. the temperature at which $m_D$ is of the size of the inverse of the  
radius of the bound state, this finding challenges the long-time accepted view 
that quarkonium remains dissociated in a medium as long as color is screened;  
in fact, the existence of the melting temperature implies that quarkonium remains dissociated 
at lower temperatures, experiencing a larger dissociation rate.

In the following, we will report about the results of \cite{Brambilla:2010vq}, where,  
in a specific range of temperatures, the spectrum and widths of quarkonia 
up to order $m\als^5$ have been computed. 
In the assumed range of parameters, the scales of the bound state ($m$, $m\als$ and $m\als^2$) 
and the thermodynamical scales ($T$ and $m_D$) fulfill a specific hierarchy that we will exploit 
either by constructing a corresponding hierarchy of EFTs or by subsequently integrating out 
different momentum regions.

\section{Scales and EFTs}
We consider quarkonium in a thermal bath, whose energy scales satisfy the following 
hierarchy:
\begin{equation}
m \gg m\als \gg \pi T \gg m\als^2 \gg m_D.
\label{hierarchy}
\end{equation}
This implies that $mg^3 \gg T \gg mg^4$ and that 
$\pi T$ is lower than $\pi T_{\rm melting}$, i.e. quarkonium exists in the plasma.
We will further assume that all these scales are larger than $\lQ$ and that 
a weak-coupling expansion is possible for all of them.
Finally, in order to produce an expression for the spectrum that is accurate up to order 
$m\als^5$, we will assume that $[m_D/(m\als^2)]^4 \ll g$. 

It is not clear if this hierarchy is realized by any of the 
quarkonium states at present heavy-ion facilities. 
Possible candidates are the bottomonium $1S$ states ($\Upsilon(1S)$, $\eta_b$) at the LHC, for which 
it may hold $m_b \approx 5 \; \hbox{GeV} \; > m_b\als \approx 1.5 \; \hbox{GeV} \; >  \pi T \approx 1 \; 
\hbox{GeV} \; >  m\als^2 \approx 0.5  \; \hbox{GeV} \; \simg  m_D$. 

It is useful to work in the so-called real-time formalism, which amounts to modifying the 
temporal integration in the partition function to include real times. As a consequence,  
the degrees of freedom double. However, the doubling of the degrees 
of freedom affects only gluon loops (at the order we are working only the gluon self energy), 
while physical heavy quarks decouple from unphysical ones \cite{Brambilla:2008cx}, so the 
price to pay is minimal. The advantage is that the real-time formalism allows a treatment 
of the quarkonium in the thermal bath very similar to the EFT framework developed 
for zero temperature \cite{Brambilla:2004jw}. Many results obtained there may, indeed, be simply 
translated here.

The EFT that follows from QCD by integrating out gluons and quarks of energy or momentum 
of order $m$ in the quark-antiquark sector is non-relativistic QCD (NRQCD) \cite{Caswell:1985ui}.
The EFT that follows from NRQCD by integrating out gluons of energy or momentum 
of order $m\als$ is potential non-relativistic QCD (pNRQCD) \cite{Pineda:1997bj}.
According to the hierarchy \eqref{hierarchy}, both these scales are larger 
than $T$, which, therefore, may be set to zero in the matching to the EFT.
As a consequence, the Lagrangians of NRQCD and pNRQCD are the same as at zero temperature.

Integrating out $T$ from pNRQCD modifies pNRQCD into hard-thermal loop (HTL) pNRQCD, 
pNRQCD$_{\rm HTL}$, \cite{Brambilla:2008cx,Vairo:2009ih}. With respect to pNRQCD, 
the pNRQCD$_{\rm HTL}$  Lagrangian gets relevant modifications in two parts.
First, the Yang--Mills Lagrangian gets an additional HTL part \cite{Braaten:1991gm}.
This, for instance, modifies the longitudinal gluon propagator in Coulomb gauge into ($k^2 \equiv {\bf k}^2$)
\begin{equation}
\frac{i}{k ^2}
\to 
\frac{i}{k^2+m_D^2\left(1-\displaystyle\frac{k_0}{2k}\ln\frac{k_0+k\pm i\eta}{k_0-k\pm i\eta}\right)},
\end{equation}
where ``$+$'' identifies the retarded and ``$-$'' the advanced propagator.
Second, the potentials get in addition to the Coulomb potential, which 
is the potential inherited from pNRQCD, a thermal part, $\delta V$.

\section{Potential, energy and decay width}
In the following, we will provide the thermal corrections to the color-singlet 
quark-antiquark potential, the thermal corrections to the spectrum and the thermal decay width, 
aiming at a precision of the order of $m\als^5$.

\begin{figure}[ht]
\makebox[0truecm]{\phantom b}
\put(-100,0){\epsfxsize=6.5truecm \epsfbox{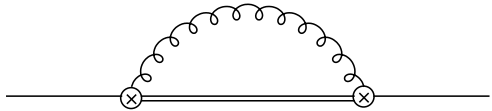}}
\label{fig1}
\caption{The single line stands for a quark-antiquark color-singlet propagator,
the double line for a quark-antiquark color-octet propagator and the circle with a cross for 
a chromoelectric dipole interaction.}
\end{figure}

\subsection{Integrating out the scale $T$}
The relevant diagram contributing to the potential is shown in Fig.~\ref{fig1}. It reads
\begin{equation}
- i g^2 \, \frac{4}{3} \, \frac{r^i}{D-1}
\mu^{4-D} \int \frac{d^Dk}{(2\pi)^D}
\frac{i}{E - h_o -k_0 +i\eta}\left[k_0^2 \, D_{ii}(k_0,k) +  k^2 \, D_{00}(k_0,k)
\right]r^i, 
\end{equation}
where $D_{\mu\nu}$ stands for the gluon propagator, $h_o = {\bf p}^2/m + \als/(6r)$ is the octet 
Hamiltonian and the loop integral has been regularized in dimensional regularization ($D = 4 + \epsilon$ 
and $\mu$ is the subtraction point). In the loop integral, we integrate over 
the momentum region $k_0\sim T$ and $k \sim T$.
Since $T\gg(E-h_o)$, we may expand 
$$
\frac{i}{E-h_o-k_0+i\eta}=
\frac{i}{-k_0+i\eta}-i\frac{E-h_o}{(-k_0+i\eta)^2}
+i\frac{(E-h_o)^2}{(-k_0+i\eta)^3}-i\frac{(E-h_o)^3}{(-k_0+i\eta)^4}+\dots\,.
$$

\begin{figure}[ht]
\makebox[0truecm]{\phantom b}
\put(-200,20){ \it a)}
\put(-190,0){\epsfxsize=6.5truecm \epsfbox{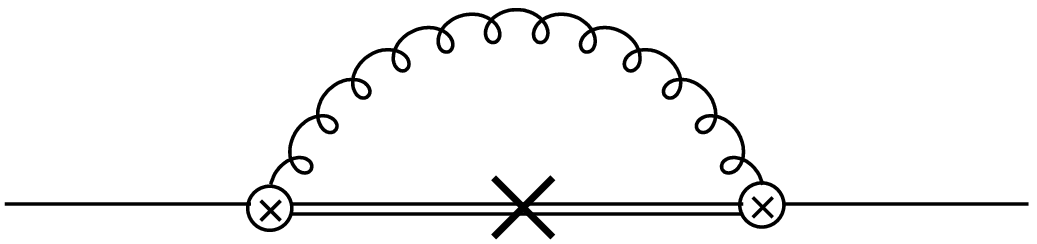}}
\put(10,20){ \it b)}
\put(20,0){\epsfxsize=6.5truecm \epsfbox{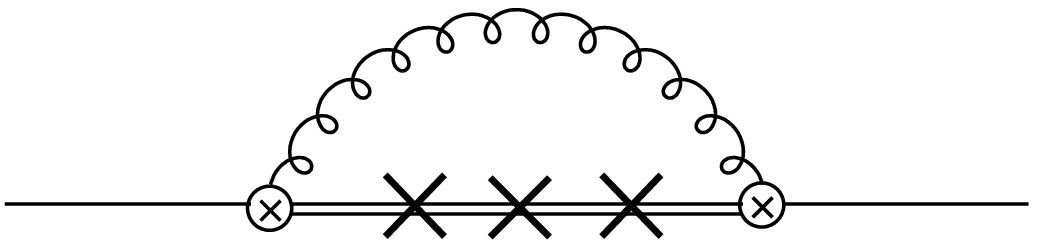}}
\put(-110,-10){$E-h_o$}
\put(-200,-60){ \it c)}
\put(-190,-80){\epsfxsize=6.5truecm \epsfbox{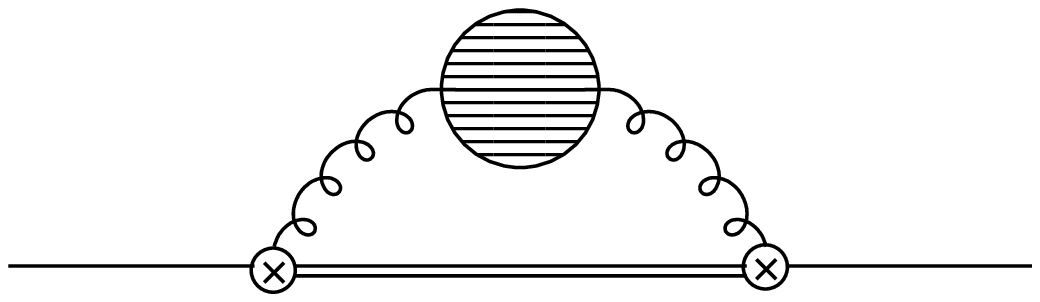}}
\label{fig2}
\caption{Detail of the different terms contributing to the potential.}
\end{figure}

The real part of the thermal correction to the color-singlet potential reads
\begin{eqnarray}
{\rm Re}~ \delta V_s(r)  &=&  \frac{4\pi}{9}  \als^2 \, r \, T^2 + \frac{8\pi}{9m} \als \,T^2
\nonumber\\
& & 
+\frac{4\als I_T}{3\pi}\left[-\frac{9}{8}\frac{\als^3}{r}-\frac{17}{3}\frac{\als^2}{mr^2}
+\frac{4}{9}\frac{\pi\als}{m^2}\delta^3({\bf r})+ \frac{\als}{m^2}\left\{\nabla^2_{\bf r},\frac{1}{r}\right\}\right]
\nonumber\\
& & - 2 \zeta(3) \frac{\als}{\pi} \, r^2 \, T \,m_D^2 + \frac{8}{3} \zeta(3) \als^2 \, r^2 \, T^3 , 
\label{realV}\\
I_T   &=&  \frac{2}{\epsilon}+\ln\frac{T^2}{\mu^2}-\gamma_E+\ln( 4\pi)-\frac{5}{3},
\\
m_D^2 &=& g^2T^2\left(1 + \frac{n_f}{6} \right),
\label{mD}
\end{eqnarray}
where the first line of Eq. \eqref{realV}, which is of order  $ g^2r^2T^3\times E/T$, 
comes from diagram {\it a)} in Fig.~\ref{fig2}, 
the second one, which is of order $g^2r^2T^3\times (E/T)^3$, comes from diagram {\it b)}, 
and the third one, which is of order $g^2r^2T^3\times ({m_D}/{T})^2$, comes from diagram {\it c)}; 
$n_f$ is the number of light quarks.

The imaginary part of the color-singlet potential, 
which comes from diagram {\it c)} in Fig.~\ref{fig2}, reads
\begin{eqnarray}
 {\rm Im}~\delta V_s(r)   \!\!\! &=&  \!\!\!
\frac{2}{9} \als \, r^2 \, T \,m_D^2\, \left( 
-\frac{2}{ \epsilon} + \gamma_E + \ln\pi 
- \ln\frac{T^2}{\mu^2} + \frac{2}{3} - 4 \ln 2 - 2 \frac{\zeta^\prime(2)}{\zeta(2)} \right)
\nonumber\\
& &  \!\!\!
+ \frac{16\pi}{9} \ln 2 \,  \als^2\, r^2 \, T^3.
\end{eqnarray}
This contribution, which may be traced back to the Landau-damping phenomenon, 
is of order $g^2r^2T^3\times \left({m_D}/{T}\right)^2$.  

Evaluating ${\rm Re}~\delta V_s(r)$ and ${\rm Im}~\delta V_s(r)$ on a quarkonium state 
with quantum numbers $n$ and $l$, we obtain the thermal correction to the energy, 
$\delta E_{n,l}^{(T)}$, and the thermal width, $\Gamma_{n,l}^{(T)}$, coming from  the scale $T$:
\begin{eqnarray}
\delta E_{n,l}^{(T)}  &=& 
\frac{2\pi}{9}\,\als^2 \,T^2 a_0 \left[3n^2-l(l+1)\right]
+\frac{8\pi}{9m} \als\, T^2
\nonumber\\
&&
+\frac{E_n { I_T}  \als^3}{3\pi}\left\{
-\frac{32}{27}\frac{\delta_{l0}}{n} 
+\frac{200}{3}\frac{1}{n (2l+1)}
-\frac{16}{3}\frac{1}{n^2}
+\frac{27}{4}
\right\}
\nonumber\\
&& 
+\left(- \zeta(3) \frac{\als}{\pi}  \, T \,m_D^2
+ \frac{4}{3} \zeta(3) \als^2 \, T^3\right) a_0^2n^2 \left[5n^2+1-3l(l+1)\right], 
\label{ET}\\
\Gamma_{n,l}^{(T)}  &=& 
\left[ - \frac{2}{9} \als T m_D^2
\left( -\frac{2}{\epsilon} + \gamma_E + \ln\pi 
- \ln\frac{T^2}{\mu^2}+ \frac{2}{3} - 4 \ln 2 - 2 \frac{\zeta^\prime(2)}{\zeta(2)} \right) \right.
\nonumber\\
&& \left.
-\frac{16\pi}{9} \ln 2 \,  \als^2\, T^3 \right] {a_0^2n^2}\left[5n^2+1-3l(l+1)\right], 
\label{GT}
\end{eqnarray}
where $\displaystyle E_n=-\frac{1}{m a_0^2 n^2} = -\frac{4 m\als^2}{9n^2}$ and $\displaystyle a_0 = \frac{3}{2m\als}$.

\subsection{Integrating out the scale $E$}
The diagram shown in Fig.~\ref{fig1} also carries contributions coming from the energy scale $E$.
They may be best evaluated in pNRQCD$_{\rm HTL}$ by integrating 
over the momentum region $k_0\sim E$ and $k \sim E$ and using HTL gluon propagators.
Since $k\sim E \ll T$, we may expand the Bose--Einstein distribution
\begin{equation}
n_{\rm B}(k)=\frac{T}{k}-\frac{1}{2}+\frac{k}{12 \, T} + \dots \;;
\label{bose}
\end{equation}
moreover, since  $k\sim E\gg m_D$, the HTL propagators can be expanded in $m_D^2/E^2\ll 1$.

The momentum region $k_0\sim E$ and $k \sim E$ is characterized by two possible 
momentum sub-regions. This can be understood by considering the integral  
\begin{equation}
\int\!\!\frac{d^{D-1}k}{(2\pi)^{D-1}}\int_0^\infty\!\!\frac{dk_0}{2\pi} 
\frac{1}{k_0^2 -k^2 - m_D^2 + i\eta}
\left(\frac{1}{E-h_o-k_0+i\eta}+\frac{1}{E-h_o+k_0+i\eta}\right). 
\end{equation}
For $k_0\sim E$ and $k \sim E$, it exhibits an off-shell sub-region, $k_0-k \sim  E$, 
and a collinear sub-region, $k_0-k \sim  m_D^2/E$.
Note that, according to \eqref{hierarchy}, the collinear scale satisfies 
$mg^4 \gg  m_D^2/E \gg mg^6$, i.e. it is smaller than $m_D$ by a factor of $m_D/E\ll 1$ 
but still larger than the non-perturbative scale $g^2 T$ by a factor $T/E \gg 1$.

The thermal correction to the energy, $\delta E_{n,l}^{(E)}$, coming from  the scale $E$, reads
\begin{equation}
\delta E_{n,l}^{(E)}  = -\frac{2\pi}{9} \als \, Tm_D^2 \, a_0^2n^2 \left[5n^2+1-3l(l+1)\right].
\label{EE}
\end{equation}
We note the complete cancellation of the vacuum contribution (which includes the Bethe logarithm) 
against the thermal contribution originating from the ``$-1/2$'' term in the expansion of the Bose--Einstein 
distribution (see Eq. \eqref{bose}).

The thermal width, $\Gamma_{n,l}^{(E)}$, coming from  the scale $E$, reads
\begin{eqnarray}
\Gamma_{n,l}^{(E)}  &=&
4\als^3T-\frac{64}{9m}\als TE_n+\frac{32}{3}\als^2T\frac{1}{mn^2a_0}
\nonumber\\
&&
+\frac{2E_n\als^3}{3}
\left\{
-\frac{32}{27}\frac{\delta_{l0}}{n} 
+\frac{200}{3}\frac{1}{n (2l+1)}
-\frac{16}{3}\frac{1}{n^2}
+\frac{27}{4}
\right\}
\nonumber\\
&&
-\frac{2}{9} \als Tm_D^2 \left(\frac{2}{ \epsilon}
+\ln\frac{E_1^2}{\mu^2}+\gamma_E-\frac{11}{3}-\ln\pi+\ln4\right) a_0^2n^2 \left[5n^2+1-3l(l+1)\right]
\nonumber\\
&&+\frac{128 Tm_D^2}{81}\frac{\als^3}{E_n^2}\,I_{n,l} \,,
\label{GE}
\end{eqnarray}
where $I_{1,0}=-0.49673$, $I_{2,0}=0.64070, \; \dots\;$~.
The leading contribution is given by the first three terms, which are of order  $\als^3 T$.
This contribution to the thermal width is generated by the possible break up of a quark-antiquark color-singlet state 
into an unbound quark-antiquark color-octet state: a process that is kinematically allowed only in a medium.
The singlet to octet break up is a different phenomenon with respect 
to the Landau damping. In the situation $E \gg m_D$, the first dominates over the second 
by a factor $(m\als^2/m_D)^2$.

\subsection{Integrating out the scale $m_D$}
The diagram shown in Fig.~\ref{fig1} also carries contributions coming from the energy scale $m_D$.
These contributions are suppressed with respect to the other terms calculated.

\section{Cancellation of divergences}
The thermal corrections to the spectrum and the thermal decay width develop divergences at the different energy scales. 
These are artifacts of the scale separations and cancel in the final (physical) results.

Concerning the thermal decay width, the divergence at the scale  $m\als^2$  in Eq. \eqref{GE}, 
which is of ultraviolet (UV) origin, cancels against the infrared (IR) 
divergence  at the scale  $T$ in Eq. \eqref{GT}.

In the spectrum, the pattern of divergences is more complicated and is summarized in Tab.~\ref{table}.
The table may be read vertically or horizontally. 
If read vertically, it shows a typical EFT cancellation mechanism:
at the scale $m\als$ we have non-thermal IR divergences in the potentials, 
these cancel against non-thermal UV divergences at the scale $m\als^2$ \cite{Brambilla:1999qa}, 
the non-thermal contribution at the scale $T$ is scaleless and vanishes in dimensional regularization; 
thermal IR divergences at the scale $T$ cancel against thermal UV divergences at the scale $m\als^2$. 
If read horizontally, it shows a cancellation mechanism that is familiar in thermal field theory: 
at the scale $T$, thermal IR divergences cancel against non-thermal IR divergences while non-thermal UV 
divergences cancel against IR divergences that appear in the potentials at the scale $m\als$; 
at the scale $m\als^2$ UV thermal divergences cancel against UV non-thermal divergences. 
The cancellation between non-thermal and thermal parts 
is possible because the latter may carry tempe\-ra\-ture independent terms (see, for instance, 
the ``$-1/2$'' term in Eq. \eqref{bose}). 
Note that both at the scales $T$ and $m\als^2$, the spectrum is finite.

\begin{center}
\begin{table}
\begin{tabular}{c|c|c}
Scale  & Vacuum & Thermal \\
\hline 
& & \\
$m\als$ &$\displaystyle \sim m\als^5\frac{1}{\epsilon_{\rm IR}}$ 
& \put(15,-5){\epsfxsize=1cm\epsfbox{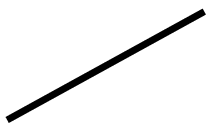}} \\
& & \\
$T$& $\displaystyle \sim m\als^5\left(\frac{1}{\epsilon_{\rm IR}} -\frac{1}{\epsilon_{\rm UV}} \right)$ 
&$\displaystyle \sim - m\als^5\frac{1}{\epsilon_{\rm IR}}$ \\
& & \\
$m\als^2$ & $\displaystyle \sim -m\als^5\frac{1}{\epsilon_{\rm UV}}$ 
& $\displaystyle \sim m\als^5\frac{1}{\epsilon_{\rm UV}}$\\
\end{tabular}
\caption{The pattern of IR and UV divergences in the quarkonium spectrum at different energy 
scales. The final result is finite.}
\label{table}
\end{table}
\end{center}

\section{Summary}
For a quarkonium state that satisfies the hierarchy specified in Eq. \eqref{hierarchy} 
and in the following discussion, 
the complete thermal contribution to the spectrum up to ${\cal O}(m\als^5)$ 
is obtained by summing Eqs. \eqref{ET} and \eqref{EE} and subtracting 
from the latter the zero-temperature part, this gives 
\begin{eqnarray}
\delta E_{n,l}^{(\mathrm{thermal})} &=& 
\frac{2\pi}{9} \als^2 \,T^2 a_0 \left[3n^2-l(l+1) + \frac{8}{3}\right] 
\nonumber\\
&&
+\frac{E_n\als^3}{3\pi}\left[\log\left(\frac{2\pi T}{E_1}\right)^2-2\gamma_E\right]
\nonumber\\
&& \hspace{4cm} \times
\left\{
-\frac{32}{27}\frac{\delta_{l0}}{n} 
+\frac{200}{3}\frac{1}{n (2l+1)}
-\frac{16}{3}\frac{1}{n^2}
+\frac{27}{4}
\right\}
\nonumber\\
&&
+\frac{128E_n\als^3}{81\pi}L_{n,l}
\nonumber\\
&&
+ a_0^2n^2 \left[5n^2+1-3l(l+1)\right]
\left\{- \left[\frac{1}{\pi} \zeta(3)+\frac{2\pi}{9} \right]   \als  \, T \,m_D^2
\right.
\nonumber\\
&& \hspace{4cm}
+ \left. 
\frac{4}{3} \zeta(3)\, \als^2 \, T^3\right\},
\end{eqnarray}
where $L_{n,l}$ are the QCD Bethe logarithms: $L_{1,0}=-81.5379$, 
$L_{2,0}=-37.6710, \; \dots\;$ \cite{Kniehl:2002br}.

For a quarkonium state that satisfies the hierarchy specified in Eq. \eqref{hierarchy} 
and in the following discussion, 
the complete thermal width up to ${\cal O}(m\als^5)$ is obtained by summing 
Eqs.~\eqref{GT} and \eqref{GE}, this gives
\begin{eqnarray}
\Gamma_{n,l}^{(\mathrm{thermal})} &=& 
\left( 4 + \frac{832}{81}\frac{1}{n^2} \right)\als^3T
\nonumber\\
&&
+\frac{2E_n\als^3}{3}
\left\{
-\frac{32}{27}\frac{\delta_{l0}}{n} 
+\frac{200}{3}\frac{1}{n (2l+1)}
-\frac{16}{3}\frac{1}{n^2}
+\frac{27}{4}
\right\}
\nonumber\\
&&
-\left[\frac{2}{9} \als T m_D^2
\left(\ln\frac{E_1^2}{T^2}+ 2\gamma_E -3 -\log 4- 2 \frac{\zeta^\prime(2)}{\zeta(2)} \right)
+\frac{16\pi}{9} \ln 2  \,  \als^2\, T^3 \right] 
\nonumber\\
&& \hspace{6cm}\times\; a_0^2n^2\left[5n^2+1-3l(l+1)\right]
\nonumber\\
&& + \frac{32}{9}\als\, Tm_D^2\,a_0^2n^4 \,I_{n,l}.
\end{eqnarray}

As a qualitative summary, we observe that, at leading order, the quarkonium masses 
increase quadratically with $T$,
which implies the same functional increase in the
energy of the leptons and photons produced in the electromagnetic decays.
Electromagnetic decays occur at short distances $\sim 1/m \ll 1/T$, 
hence the standard NRQCD factorization formulas hold. At leading order, 
all the temperature dependence is encoded in the wave function
at the origin. The leading temperature correction to it can be read from the 
potential and is of order $n^4 T^2/(m^2\als)$. 
Hence, a  quadratic dependence on the temperature should
be observed in the frequency of produced leptons or photons.
Finally, at leading order, a decay width linear with temperature is developed.
The mechanism underlying this decay width is the color-singlet to color-octet 
thermal break-up, which implies a tendency of the quarkonium to decay 
into a continuum of color-octet states.

\begin{theacknowledgments}
We acknowledge financial support from the RTN Flavianet MRTN-CT-2006-035482 (EU) and 
from the DFG cluster of excellence ``Origin and structure of the universe'' 
(http://www.universe-cluster.de). 
\end{theacknowledgments}

\bibliographystyle{aipprocl}

\end{document}